# Ordering of Interstitial Iron Atoms and Local Structural Distortion Induced by Iron Polycomplex in $Fe_{1+y}Te_{1-x}Se_x$ as Seen via Transmission Electron Microscopy

Xiao-Ping Ma,[1, 2, 3] Lu Zhang,[1, 4] Wen-Tao Wang,[1, 4] Jing-Zhe Nie,[1, 4] Huan-Fang Tian,[1, 4] Shi-Long Wu,[1, 4] Shuai-Shuai Sun,[1, 4] Tian-Long Xia,[2, 3, 5, 6] Jun Li,[1, 4, *] Jian-Qi Li,[1, 4, 7, *] and Huai-Xin Yang[1, 4, *]

1 Beijing National Laboratory for Condensed Matter Physics and Institute of Physics, Chinese Academy of Sciences, Beijing 100190, China
2 Department of Physics, Renmin University of China, Beijing 100872, China
3 Beijing Key Laboratory of Opto-electronic Functional Materials & Micro-nano Devices, Renmin University of China, Beijing 100872, China
4 School of Physical Sciences, University of Chinese Academy of Sciences, Beijing 100190, China
5 Key Laboratory of Quantum State Construction and Manipulation (Ministry of Education), Renmin University of China, Beijing, 100872, China
6 Laboratory for Neutron Scattering, Renmin University of China, Beijing 100872, China
7 Songshan Lake Materials Laboratory, Dongguan, Guangdong, 523808, People's Republic of China

**Abstract**

Employing aberration-corrected scanning transmission electron microscopy (STEM), we meticulously investigated the intrinsic chemical heterogeneity of $Fe_{1+y}Te$, $Fe_{1+y}Te_{0.8}Se_{0.2}$, and $Fe_{1+y}Te_{0.5}Se_{0.5}$. Comprehensive analysis reveals the presence of interstitial iron atoms ($Fe_{int}$) across all samples, predominantly occupying the 2c site of the $P4/nmm$ space group. Moreover, a superstructure phase characterized by a wave vector $\boldsymbol{q} = 2/5\boldsymbol{a} + 1/2\boldsymbol{c}$, originating from the ordering of $Fe_{int}$, is distinctly observable in the parent compound $Fe_{1+y}Te$. In this scenario, the $Fe_{int}$ atoms interact with adjacent Fe atoms, forming iron polycomplex and leading to an evident distortion of the $FeTe_4$ tetrahedral. Experimental results further demonstrate effective suppression of $Fe_{int}$ concentration and ordering through appropriate Se substitution; notably, $Fe_{1+y}Te_{0.5}Se_{0.5}$ manifests the lowest concentration of $Fe_{int}$ atoms. Our findings additionally indicate that Se substitution is random, and nanoscale phase separation induced by Te/Se chemical heterogeneity is commonly observed within $Fe_{1+y}Te_{1-x}Se_x$ ($0 \leq x \leq 1$) crystals.

## 1. Introduction

The discovery of iron-based superconductors has captured significant attention within the condensed matter physics community[1, 2]. Several structural families of iron-based superconductors have been identified, providing a new platform for investigating unconventional high-temperature superconductivity[3-19]. Among these, the 11-type iron chalcogenides (Fe$Ch$) superconductors are particularly noteworthy due to their simple crystal structure, characterized by Fe$Ch$ layers stacked along the $c$-axis without intermediate spacer layers[19]. The initial superconducting transition temperature ($T_c$) of FeSe was ~ 8 K[19], which was enhanced to around 14.5 K through appropriate Te substitution[20-23], and further elevated to 37 K under high pressure[24-26]. Furthermore, angle-resolved photoemission spectroscopy (ARPES) revealed a substantial superconducting gap ~19 meV in a one unit-cell FeSe thin film grown on $SrTiO_3$, suggesting a potential $T_c$ as high as 65 K[27], and a sign of $T_c$ even above 100 K was also reported[28]. The other end-member of this family, FeTe, is not superconducting but instead exhibits an antiferromagnetic ordering coupled with a first-order structural phase transition at approximately 70 K[29, 30]. Superconductivity was further induced by substituting S or Se at the Te site[20-23, 31, 32], or by oxygen annealing of FeTe thin films[33-36].

Contrary to their simplistic crystalline structure, FeTe$_{1-x}$Se$_x$ ($0 \leq x \leq 1$) compounds tend to exhibit significant nonstoichiometric disordering. Two major types of chemical inhomogeneity are present: sample-dependent Fe non-stoichiometries[20-23, 29, 30, 37, 38], and non-uniform spatial distribution of Te/Se[39-41]. These chemical inhomogeneities are commonly observed in FeTe$_{1-x}$Se$_x$ ($0 \leq x \leq 1$) materials and profoundly impact the delicate interplay between crystalline structure and physical properties[20-23, 29, 30, 38-41]. Previous investigations have revealed that the crystal growth process inevitably introduces excess Fe contents in FeTe$_{1-x}$Se$_x$ ($0 \leq x \leq 1$)[20, 21, 37-40], resulting in a non-stoichiometric composition denoted as Fe$_{1+y}$Te$_{1-x}$Se$_x$ ($0 \leq x \leq 1$), where y represents excess interstitial Fe (Fe$_{int}$) at interstitial sites, as shown in Figure 1(a). Fe$_{1+y}$Te samples with y ranging from 0.04 to 0.18 have been successfully grown and characterized, and results demonstrated that excess Fe$_{int}$ significantly affects its structural and magnetic properties; when y > 0.12, a transition from tetragonal to orthorhombic symmetry occurs, accompanied by an incommensurate magnetic phase at low temperatures, whereas monoclinic symmetry and commensurate magnetic structure were observed for samples with lower levels of iron[30, 42].

Density functional theory (DFT) calculations, based on a 2×2 supercell of tetragonal FeTe with one excess Fe$_{int}$ on Fe$_{1.125}$Te, have revealed that the Fe$_{int}$ with a valence near Fe$^+$, acts as an electron dopant, introducing additional charge carriers into the 11 system[43]. Furthermore, the Fe$_{int}$ is strongly magnetic, contributing local magnetic moments that interact with adjacent Fe layers. The magnetic moment from Fe$_{int}$ acts as a pair breaker and localizes the charge carriers[37, 43]. However, the presence of Fe$_{int}$ as a magnetic dopant on interstitial sites hampers the study of intrinsic superconductivity and topological properties. To address this challenge, considerable efforts have been made to remove excess Fe$_{int}$ and enhance superconductivity. Methods such as post-annealing treatment[44-47] or immersion in alcohol and nitric acid[48, 49] have been successfully employed, enabling researchers to explore the intrinsic properties of FeTe$_{1-x}$Se$_x$ ($0 \leq x \leq 1$) single crystals.

In the case of Te/Se inhomogeneity, scanning tunneling microscopy (STM) studies have demonstrated a nonuniform distribution of Te and Se ions, resulting in phase-segregated regions ranging from sub-micrometer to tens of nanometers in the *ab* plane.[40, 41] Conversely, Yamada and co-workers proposed that Se and Te do not actually share the same crystallographic site but rather organize into an ordered arrangement resembling a supercell structure[50]. Recent advancements have illuminated the presence of topological surface superconductivity and Majorana-bound states in FeSe$_{0.45}$Te$_{0.55}$ single crystals, as evidenced by angle-resolved photoemission spectroscopy (ARPES) and scanning tunneling spectroscopy (STS) measurements[51, 52]. Nerveless, it is noteworthy that not all vortex cores exhibit Majorana zero modes. Kong *et al.* observed that the Majorana zero modes and ordinary vortices coexist in the same field, typically grouped together within the same class. This phenomenon is likely attributed to the intrinsic chemical inhomogeneity of Se/Te within Fe$_{1+y}$Te$_{1-x}$Se$_x$ compounds ($0 \leq x \leq 1$)[52].

Consequently, despite substantial efforts directed towards Fe$_{1+y}$Te$_{1-x}$Se$_x$ compounds, a comprehensive understanding of their microstructural characterization remains a critical priority. The scanning transmission electron microscopy (STEM) technique, with sub-angstrom spatial resolution, has been extensively applied for structural characterizations, particularly for crystals exhibiting local lattice variations such as occupation or displacive modulation. In this study, we present the spatial correlation of intrinsic chemical inhomogeneities involving Fe$_{int}$, Te, and Se across three samples with nominal compositions (Fe$_{1+y}$Te, Fe$_{1+y}$Te$_{0.8}$Se$_{0.2}$, and Fe$_{1+y}$Te$_{0.5}$Se$_{0.5}$) using STEM techniques. Atomically resolved high-angle annular dark field (HAADF) images clearly illustrate that the interstitial Fe$_{int}$ atoms are highly localized at the 2*c* Wyckoff sites of the space group *P4/nmm* (as shown in Figure 1(a)), and form a modulation structure with a vector $q = 2/5a + 1/2c$ in Fe$_{1+y}$Te.

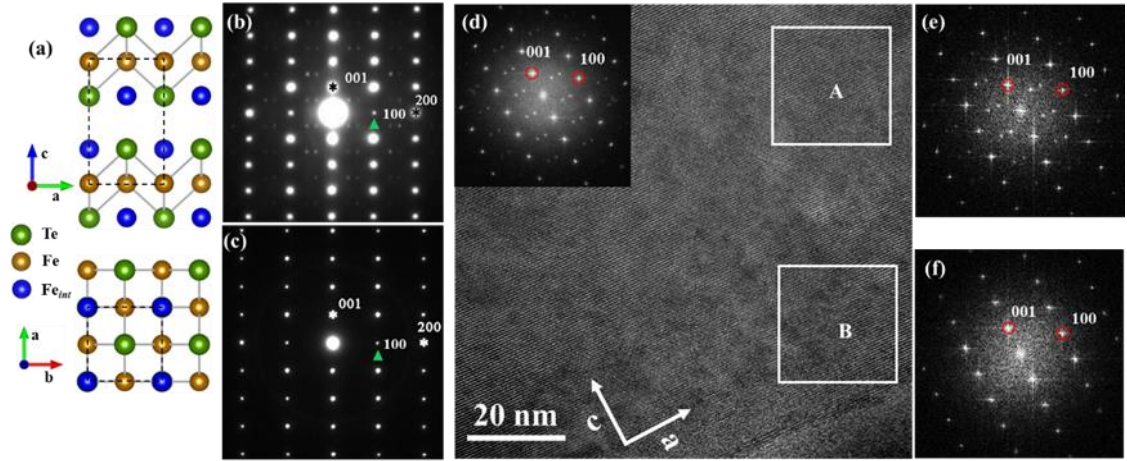

Figure 1. Atomic structure model and HRTEM images of $Fe_{1+y}Te$. (a) Atomic structure of $Fe_{1+y}Te$, where 'y' represents excess amount of $Fe_{int}$ at interstitial sites. The Te, Fe, and $Fe_{int}$ atoms are represented by light green, orange, and blue spheres, respectively. (b) and (c) show two typical selected-area electron diffraction (SAED) patterns taken along the [010] zone axis from $Fe_{1+y}Te$ single crystals. (d) A typical low-magnification HRTEM image and its corresponding FFT pattern (in the upper-left inset) of an $Fe_{1+y}Te$ single crystal. (e-f) FFT images from the two areas A and B, marked by white rectangles in (d), respectively.

Moreover, interstitial $Fe_{int}$ atoms interact with neighboring Fe atoms to form iron polycomplex, resulting in distortion of the tetrahedral $FeTe_4$ structure. Additionally, Se doping effectively reduces the content of $Fe_{int}$ atoms and disrupts their ordered distribution. Among the $Fe_{1+y}Te$, $Fe_{1+y}Te_{0.8}Se_{0.2}$, and $Fe_{1+y}Te_{0.5}Se_{0.5}$ samples, $Fe_{1+y}Te$ contains the highest amount of $Fe_{int}$ atoms, followed by $Fe_{1+y}Te_{0.8}Se_{0.2}$, with the least amount in $Fe_{1+y}Te_{0.5}Se_{0.5}$. Furthermore, several nanoscale Te- and Se-enriched regions are observed in the $Fe_{1+y}Te_{0.8}Se_{0.2}$ and $Fe_{1+y}Te_{0.5}Se_{0.5}$ samples, indicating local chemical inhomogeneities of Te and Se.

## 2. Results and discussion

Single crystals of $Fe_{1+y}Te_{1-x}Se_x$ ($0 \leq x \leq 1$) samples were synthesized using self-flux method, with complete synthetic protocols detailed in the Supplementary information (SI). Single crystal X-ray diffraction (XRD) of $Fe_{1+y}Te$, $Fe_{1+y}Te_{0.8}Se_{0.2}$, and $Fe_{1+y}Te_{0.5}Se_{0.5}$ were performed at room temperature, with the resulting XRD patterns shown in Figure S1 (SI). Only the (00*l*) peaks were observed, indicating that the crystallographic *c*-axis is perfectly perpendicular to the plane of the single crystal. The chemical composition analysis based on energy-dispersive spectroscopy (EDS), as illustrated in Figure S2 (SI), revealed average atomic ratios of Fe/Te (Se) = 1.09, 1.05, and 1.03 for $Fe_{1+y}Te$, $Fe_{1+y}Te_{0.8}Se_{0.2}$, and $Fe_{1+y}Te_{0.5}Se_{0.5}$, respectively, indicating a slight excess of Fe in each sample. Additionally, the actual Se content was found to be lower than the nominal content in $Fe_{1+y}Te_{0.8}Se_{0.2}$ and $Fe_{1+y}Te_{0.5}Se_{0.5}$ samples, likely due to the evaporation of Se during the crystal growth process. The non-superconducting $Fe_{1+y}Te$ sample exhibits an antiferromagnetic ordering coupled with a first-order structural phase transition at about 70 K (as detailed in Figure S3, SI). Superconductivity was further induced by substituting Se at the Te site in $Fe_{1+y}Te_{0.8}Se_{0.2}$ and $Fe_{1+y}Te_{0.5}Se_{0.5}$ samples, as detailed in Figure S3, (SI).

Detailed crystal-structure investigations were carried out on all three types of $Fe_{1+y}Te_{1-x}Se_x$ ($0 \leq x \leq 1$) samples using selected-area electron diffraction (SAED) and aberration-corrected STEM. Two typical SAED patterns taken along the [010] zone axis can be identified in $Fe_{1+y}Te$ single crystals, as shown in Figure 1(b) and 1(c), where the main diffraction spots with relatively strong intensities can be well indexed by standard anti-PbO type structure of the space group *P4/nmm*. However, it is noteworthy that the diffraction pattern in

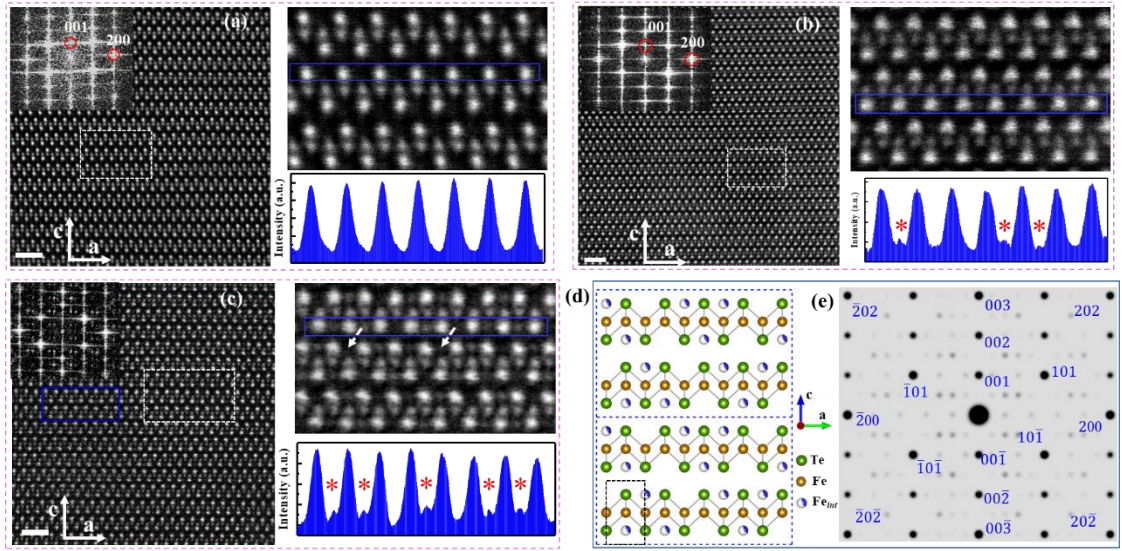

Figure 2. Three representative HAADF images of $Fe_{1+y}Te$ taken along the [010] direction. (a) A HAADF image displays pristine FeTe composition without interstitial $Fe_{int}$ atoms. (b) and (c) HAADF images revealing regions with $Fe_{int}$ atoms that are partially ccupied. The intensity line profile along the marked Te layer is presented, with $Fe_{int}$ atoms denoted by stars. (d-e) The modulated structural model with $Fe_{int}$ atoms and the corresponding simulated diffraction pattern, where Te, Fe, and $Fe_{int}$ atoms are presented by light green, orange, and blue spheres, respectively. The outer dotted rectangle represents the unit cell of the new atomic model, while the black dotted rectangle signifies the unit cell of the original anti-PbO type FeTe crystal structure. The scale bar represents 1 nm.

Figure 1(b) contains a series of weak satellite spots. The satellite spots correspond to a structural modulation with a wave vector $q = 2/5a + 1/2c$. Additionally, the presence of forbidden reflections from (*l*00) planes, where *l* = odd integers (marked by green triangles in Figure 1(b) and 1(c)), could be associated with dynamical diffraction effects. A high-resolution TEM (HRTEM) image with a large field of view recorded along the [010] direction, and the corresponding fast Fourier transform (FFT) from the whole image (left-upper inset) are shown in Figure 1(d). Two different FFT images from two areas of the same HRTEM image, indicated by white rectangles in Figure 1(e) and 1(f), reveal the existence of phase inhomogeneity.

To achieve a comprehensive understanding of the detailed structure of $Fe_{1+y}Te$ and explore the origin of superlattice spots, an extensive collection of HAADF-STEM images was acquired and categorized into three distinct types, as shown in Figure 2. It is well-established that the intensities of the atomic columns in HAADF images are proportional to the atomic number Z of the constituent elements (approximately $Z^{1.7}$)[53, 54]. Consequently, the relatively brighter dots in the present images correspond to the projections of the Te columns, and the less bright Fe column chains are located between the Te layers, which is consistent with the standard anti-PbO structure in the [010] projection. Figure 2(a) illustrates a clean FeTe region without interstitial $Fe_{int}$ atoms. In contrast, the *2c* Wyckoff sites can exhibit partial occupancy by $Fe_{int}$ atoms, as demonstrated in Figure 2(b) and 2(c) respectively. The intensity line profiles depicted in the magnified insets further corroborate the presence of $Fe_{int}$ atoms. Notably, the intensities attributed to $Fe_{int}$ are significantly lower than those corresponding to Fe at normal sites, indicating that $Fe_{int}$ atoms do not extend throughout the entire thickness of the sample. This observation is consistent with the EDS analysis, which indicates that only a very small amount of excess $Fe_{int}$ was found (see Figure S2, SI). Furthermore, the FFT patterns shown as insets in Figure 2(b) and 2(c) exhibit fundamental differences, indicating distinct arrangements of $Fe_{int}$ atoms.

Based on experimental observations, a new atomic model was constructed, as illustrated in Figure 2(d),

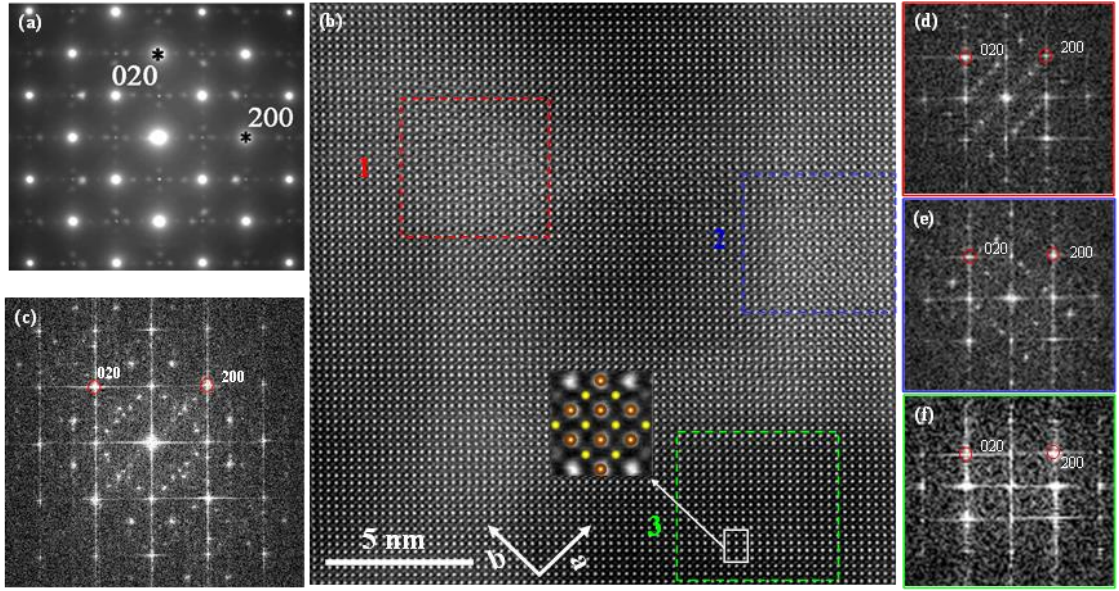

Figure 3. SEAD and HAADF images of $Fe_{1+y}Te$ taken along the [001] axis. (a) An electron diffraction pattern of $Fe_{1+y}Te$ taken along the [001] zone axis. (b) and (c) A HAADF image and the corresponding FFT patterns of the $Fe_{1+y}Te$, respectively. A typical high-magnification HAADF image recorded in a region with no modulation, with a structure consistent with the standard anti-PbO lattice shown in the inset of Figure3(b). (d-f) show three FFT patterns obtained by performing FFT on the three areas 1-3, indicated by red, blue, and green rectangles, respectively.

delineating the ordered arrangements of $Fe_{int}$ atoms. Compared to the original anti-PbO type unit cell of FeTe, the supercell expanded by a factor of 5 along the *a*-axis and 2 along the *c*-axis due to the ordering of the $Fe_{int}$ atoms. The simulated diffraction pattern based on the new atomic model accurately reproduces the superlattice spots observed in experiments, as shown in Figure 2(e). The complexity of the $Fe_{int}$ inhomogeneity can be readily introduced during crystal growth and influenced by experimental conditions, which may lead to inconsistent measurement results from different research groups. The presence of interstitial $Fe_{int}$ states significantly affects the electronic and magnetic properties of the samples[43]. Notably, an evident local distortion has been observed in the interstitial $Fe_{int}$ ordered region, as highlighted by white arrows in the inset of Figure 2(c) and Figure S4 (SI). This suggests that the presence of interstitial $Fe_{int}$ atoms triggers distinct structural distortions in the neighboring Fe atoms. A comprehensive discussion on structural distortion is provided in the subsequent section.

The $Fe_{1+y}Te$ samples underwent further characterization along the [001] direction. Figure 3(a) presents a typical [001] direction SAED pattern, where the main diffraction spots with relatively strong intensities are identified as the basic Bragg reflections indexed with the standard tetragonal anti-PbO type structure (*P4/nmm*). Each main spot is accompanied by a series of satellite reflections characterized by a modulation wave vector $q_1 = 2/5a$. Figure 3(b) displays an atomic-resolved HAADF image of $Fe_{1+y}Te$ exhibiting heterogeneous contrast. FFT patterns derived from the entire image and three distinct areas, delineated by red, blue, and green rectangles, are shown in Figure 3(c-f), respectively. Structural modulation is observed in the bright regions, as indicated in Figure 3(d) and 3(e), with the modulation wave vectors being perpendicular to each other, one points to (002) direction and the other points to (020) direction. In contrast, the modulation is absent in the dark regions as demonstrated in Figure 3(f). A higher magnification HAADF image of the modulation-free region is presented in the inset of Figure 3(b), corroborating the standard anti-PbO lattice structure (P*4/nmm*).

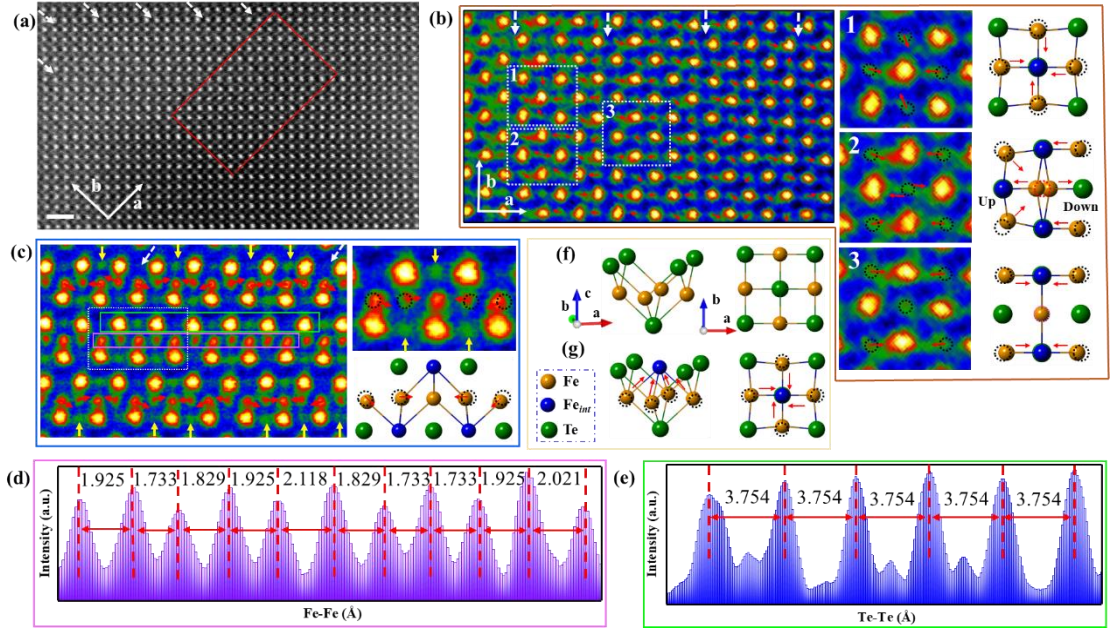

Figure 4. Local structural distortions in the Fe$_{1+y}$Te sample. (a) A HAADF image recorded in a region with modulation, where dark, gully-like stripes are clearly visible, as indicated by white arrows. The lower-right inset shows the corresponding FFT image. (b) An enlarged and rendered image taken from a portion of Figure 4(a), the inserted red arrows on the Figure 4(b) illustrate the displacement of normal Fe atoms. (c) An enlarged and rendered image from the marked green rectangle in Figure 2(c), is displayed in Figure 4(c). Yellow arrows indicate Fe$_{int}$ atoms, and white arrows indicate the direction of the normal Fe shift. (d) and (e) show the line profiles of Fe and Te columns from Figure4(c). (f-g) Comparison of the structures of FeTe with Fe$_{1+y}$Te, as shown in Figure 4(f) and Figure 4(g), reveals that Fe$_{int}$ interacts with normal Fe, formatting a contracted tetrahedral iron polycomplex. According to the new atomic model, the simulated atomic structure accurately reproduces the experimental results, as shown in Figure 4(h) and Figure 4(l). The scale bar represents 1 nm.

Figure 4(a) presents a magnified HAADF image of area 1 in Figure 3(b), clearly showing structural modulation along *a*-axis. A periodic variation in the intraplane distance is highlighted by white arrows. Detailed analysis reveals that the structural modulation originates from local distortion, where certain Fe atoms periodically deviate from their high symmetry positions in the FeTe layer. Figure 4(b) presents an enlarged view of the modulated structure extracted from Figure 4(a) with 45° rotation for clarity, showing evident displacement of Fe atoms as indicated by red arrows. A clearer view of the interaction between the Fe$_{int}$ atoms and Fe atoms in the FeTe layer was achieved in HAADF image taken along the [010] direction, as shown in Figure 4(c), which is extracted from the marked green rectangle in Figure 2(c). Figure 4(d) and Figure S4(d) (SI) display the line profile marked in Figure 4(c), providing further evidence of the periodic displacements of Fe atoms and the associated structural modulation. The magnified HAADF image shown in the inset of Figure 4(c), extracted from the region indicated by the white dash rectangle in Figure 4(c), reveals interactions between the Fe$_{int}$ atoms and the surrounding Fe atoms, resulting in a reduced bond length between Fe$_{int}$-Fe and a decreased distance between anion Te and Fe plane (anion height). Notably, the line profile in Figure 4(e) and Figure S4(e) (SI) show no significant atomic displacement within the Te columns, indicating that the Fe$_{int}$ atoms exert a much greater influence on the Fe atoms than on the Te atoms.

Based on experimental data, a comparative analysis of the structures of FeTe and Fe$_{1+y}$Te is displayed in Figure 4(f) and 4(g). This illustrates the interaction between the Fe$_{int}$ atoms and Fe atoms, resulting in the formation of iron polycomplex. This interaction induces a dual effect within the FeTe$_4$ tetrahedra: it not only leads to distortion of the bond angles within the FeTe$_4$ tetrahedra, but also causes a reduction of the height of chalcogen species from the Fe plane within these tetrahedra. The inset of Figure 4(b) illustrates three typical local distortions with different contrast feature of iron columns, which agree with the modulated structural model of Figure 2(d). In region 1, a typical iron polycomplex is demonstrated, where the Fe$_{int}$ atom is located at the center of the nearest four Fe atoms, exerting an equal attractive force on these surrounding four Fe atoms, thereby forming a contracted tetrahedral iron polycomplex. In region 2, Fe$_{int}$ atoms occupy different layers ("Up" indicates Fe$_{int}$ atoms in the upper layer, "Down" indicates Fe$_{int}$ atom in the lower layer), leading to opposing directional forces on the central Fe atoms and causing them to shift in opposite directions, resulting in a blurring of the Fe atom column at the center position. In region 3, the Fe$_{int}$ atoms exert attractive forces on the Fe atoms on the left and right sides, causing the Fe atoms on both sides to converge towards the Fe$_{int}$ atoms, while the Fe atom at the central position remains displaced due to the balanced forces. It is noteworthy that in most of the Fe$_{1+y}$Te samples, the distribution of Fe$_{int}$ atoms is complex. Different Fe$_{int}$ occupying states could result in intricate interactions between the lattice Fe atoms and nearest neighbors Fe$_{int}$ atoms, giving rise to differently distorted local structure.

Our structural investigations have uncovered the widespread presence of Fe$_{int}$ atoms in Fe$_{1+y}$Te, which interact with surrounding Fe atoms, resulting in the formation of a distorted FeTe$_4$ tetrahedron and a decreased distance in the anion height. Previous studies have revealed that Fe$_{int}$ atoms possess strong magnetic properties[37, 43]. Our results further suggest that this strong magnetism does not merely originate from Fe$_{int}$ atoms but is more likely the result of iron polycomplex formed by the interactions between Fe$_{int}$ atoms and their neighboring iron atoms. Consequently, iron polycomplex not only act as pair breaker and localize the charge carriers, thereby destroying the superconductivity, but also induce local changes in crystallographic symmetry and lattice structure. Such distorted structure subsequently has a profound impact on the properties of the Fe$_{1+y}$Te$_{1-x}$Se$_x$ system. A remarkable feature in Fe-based superconductors is the strong correlation between the local crystal structure and superconductivity. The Optimal values of Tc are obtained when the Fe(Ch)$_4$-tetrahedrons form a regular shape and the anion height to approximately 1.38 Å[55, 56]. Additionally, Choi et al. utilized DFT calculations and discovered that the stability of magnetic phases is very sensitive to the height of chalcogen species from the Fe plane. While FeTe with optimized Te height exhibits the double-stripe ($\pi$, 0) magnetic ordering, single-stripe ($\pi$, $\pi$)ordering becomes the ground state when Te is lowered below a critical height by, for example Se doping[57]. Moreover, the nontrivial topology is mainly controlled by the Te(Se) height. Adjusting the anion height, which can be achieved by varying the lattice constants and x in FeTe$_{1-x}$Se$_x$, can drive a topological phase transition[58].

Previous studies have suggested that the amount of Fe$_{int}$ can be effectively reduced by substituting Te with the smaller anion Se/S, and this reduction of Fe$_{int}$ is beneficial for enhancing superconductivity[20, 32]. Consequently, two representative samples, Fe$_{1+y}$Te$_{0.8}$Se$_{0.2}$ (displaying a small superconducting volume fraction) and Fe$_{1+y}$Te$_{0.5}$Se$_{0.5}$ (exhibiting a large superconducting volume fraction), were synthesized and characterized (see Figure S3(c) and S3(d), SI). As shown in Figure 5, HAADF images of Fe$_{1+y}$Te$_{0.8}$Se$_{0.2}$ and Fe$_{1+y}$Te$_{0.5}$Se$_{0.5}$ samples, taken along the [010] direction, are presented. it is evident that the concentration of Fe$_{int}$ in the superconducting sample Fe$_{1+y}$Te$_{0.8}$Se$_{0.2}$ is significantly lower than that in the non-superconducting Fe$_{1+y}$Te sample, and Fe$_{1+y}$Te$_{0.5}$Se$_{0.5}$ contains the least amount of Fe$_{int}$, which concurs with the previous reports[20]. Moreover, the absence of superlattice spots in the FFTs and SAED patterns (see Figure S5, SI), indicates that Se doping effectively suppresses the ordering of Fe$_{int}$ atoms.

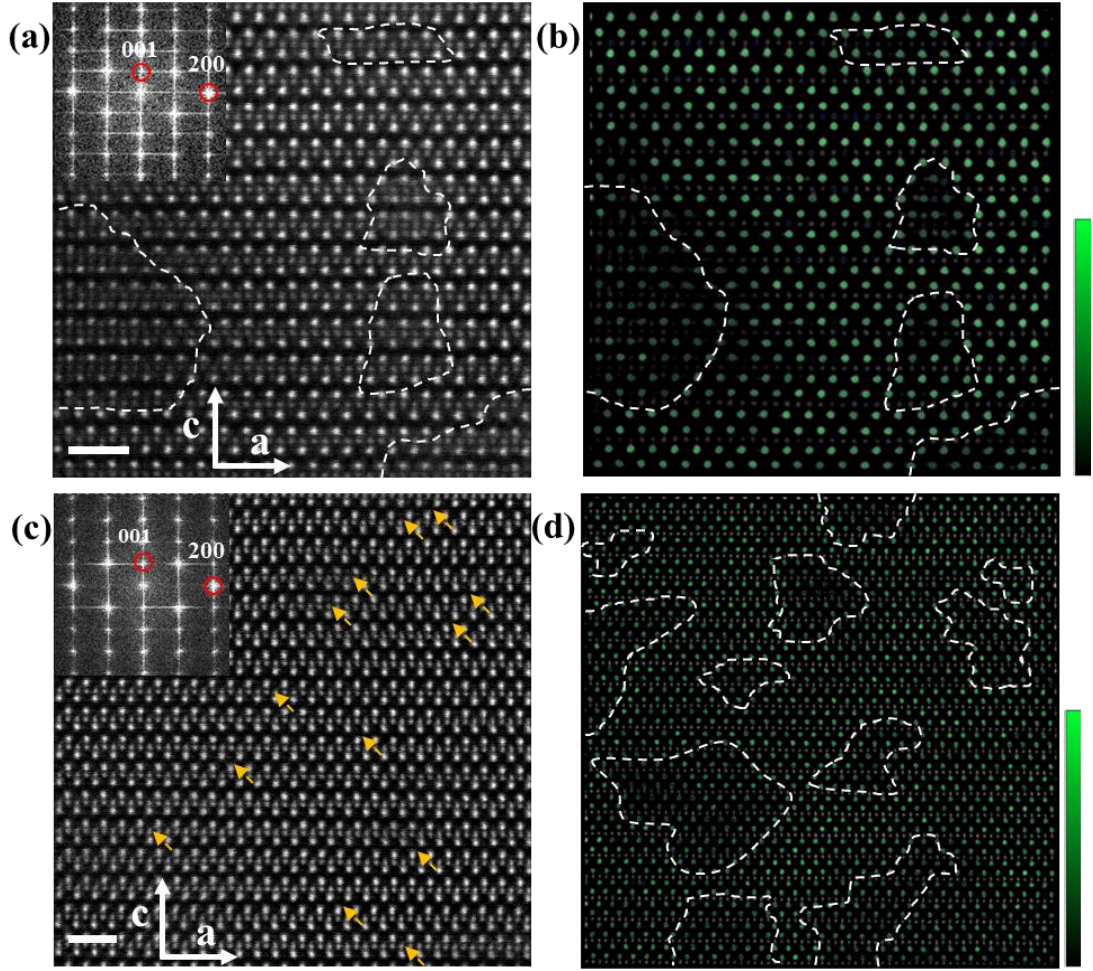

Figure 5. HAADF images of $Fe_{1+y}Te_{0.8}Se_{0.2}$ and $Fe_{1+y}Te_{0.5}Se_{0.5}$ samples taken along the [010] zone axis. (a, c) Te and Se atoms demonstrate a propensity for forming Te/ Se-rich domains. Yellow arrows indicate $Fe_{int}$ atoms. The FFT images in the upper left inset indicate that the ordered distribution of $Fe_{int}$ atoms observed in $Fe_{1+y}Te$ sample is disrupted in the Se-doped $Fe_{1+y}Te_{0.8}Se_{0.2}$ and $Fe_{1+y}Te_{0.5}Se_{0.5}$ samples. (b, d) Corresponding atomic intensities of Te/Se columns identified by 2D Gaussian fitting. The dark regions enclosed by white dashed lines indicate areas of Se-rich domains. The scale bar represents 1 nm.

Further analysis of the HAADF images reveals non-uniform contrast in the chalcogen column of these two samples. Given that the brightness of atomic columns in HAADF images is proportional to the atomic number, we identified the bright atom as Te and the dark ones as Se at the chalcogen sites.[53, 54] This indicates that Te and Se atoms prefer forming Te/Se-rich domains rather than occupying distinct lattice sites with a superlattice structure in the $Fe_{1+y}Te_{0.8}Se_{0.2}$ and $Fe_{1+y}Te_{0.5}Se_{0.5}$ samples[50]. For better visualization, we extracted the relative intensity of Te/Se layer atomic columns using two-dimension Gaussian fitting, which are represented with varying shades of green in Figures 5(b) and 5(d), respectively. We noticed that the size of these domains is relatively small, generally only a few nanometers (~2 nm, more examples see Figure S6 and S7, SI), which is less than the superconducting coherence length of $Fe_{1+y}Te_{1-x}Se_x$ ($0 \leq x \leq 1$), $\xi_c$~3 nm[59]. Similar chemical inhomogeneity in the Te/Se distribution was also observed along the [001] direction, as illustrated in Figure S8 (SI). The influence of Te/Se inhomogeneity on electronic, superconducting, and magnetic properties has been widely discussed in previous literatures[40, 41, 57, 58, 60]. For example, Yue *et al.* have observed a systematic suppression of superconductivity with decrease flake thickness, attributing this to nanoscale phase separation due to the inhomogeneous distribution of Te/Se[40]. Overall, the presence of nanoscale inhomogeneity of Te/Se leads to more inhomogeneous properties in $Fe_{1+y}Te_{1-x}Se_x$ ($0 \leq x \leq 1$) system. Additionally, Zunger *et al.* also proposed the concept of a polymorphous network, revealing that FeSe may exhibit various symmetry-breaking microstructures at the local atomic level[61]. The presence and interactions of these local microstructures significantly influence the electronic structure of FeSe, rather than it being solely determined by the average crystal structure. Ultimately, these complex local microstructures reduce the overall symmetry of FeSe's electronic structure.

## 3. Conclusion

In conclusion, we characterized a comprehensive investigation of the microstructure, with a particular focus on intrinsic chemical inhomogeneity, in three representative samples with nominal stoichiometric compositions: $Fe_{1+y}Te$, $Fe_{1+y}Te_{0.8}Se_{0.2}$, and $Fe_{1+y}Te_{0.5}Se_{0.5}$. Our analyses, particularly utilizing HAADF-STEM, clearly reveal that $Fe_{int}$ atoms are situated at the Wyckoff position $2c$ site in the chalcogen layer of the space group P$4/nmm$ in all three samples. In $Fe_{1+y}Te$, a wave vector $\boldsymbol{q} = 2/5\boldsymbol{a} + 1/2\boldsymbol{c}$ caused by $Fe_{int}$ ordering has been observed. Our experiment reveals that the $Fe_{int}$ atoms interact with interact with nearest neighbor lattice Fe atoms, leading to an evident distortion of the local structure. Then, experimental findings demonstrate effective suppression of $Fe_{int}$ concentration and ordering via suitable Se substitution; notably, $Fe_{1+y}Te_{0.5}Se_{0.5}$ exhibits the lowest concentration of $Fe_{int}$ atoms. Additionally, our results also suggest that Se substitution is random and nanoscale phase separation induced by Te/Se chemical heterogeneity, with Te/Se-rich domains of a few nanometers (~2 nm) frequently observed within $Fe_{1+y}Te_{1-x}Se_x$ ($0 \leq x \leq 1$) crystals. These findings suggest the existence of nanoscale phase separation or chemical inhomogeneity throughout the $Fe_{1+y}Te_{1-x}Se_x$ ($0 \leq x \leq 1$) systems. It is plausible that the existence of these nanoscale phases or chemical inhomogeneity contributes to the difficulties faced by different experimental groups in reproducing identical $Fe_{1+y}Te_{1-x}Se_x$ ($0 \leq x \leq 1$) samples, even under identical experimental conditions. Overall, this work provides atomic-scale insight into the fundamental crystal chemistry, offering a rational pathway for understanding the intricate physical properties observed in the $Fe_{1+y}Te_{1-x}Se_x$ ($0 \leq x \leq 1$) system.

## 4. Experimental Section

The single crystals of $Fe_{1+y}Te_{1-x}Se_x$ ($0 \leq x \leq 1$) samples were synthesized using a self-flux method. The starting materials of high-purity elemental Fe, Te and/ or Se powders were mixed by a molar ratio, then placed into alumina crucibles and sealed in evacuated quartz tubes. The sealed tubes were slowly heated up to 1070 °C at a rate of 2 °C/$min$ and kept at this temperature for 36 hours, followed by a slow cooling process to 710 °C at a rate of 2 °C/$h$, where the temperature was held for an additional and 36 hours before the furnace was shut down. Finally, shiny plate-like single crystals with dimensions of ~ 2 mm×2 mm×1 mm was readily obtained using this method.

X-ray diffraction (XRD) measurements were carried out with a Bruker AXS D8 Advanced diffractometer equipped with Cu $K\alpha$ radiation at 40 kV and 40 mA with 2θ ranging from 5° to 80°, a step with of 0.02°, and a counting time of 2 s/step. Chemical composition and microstructure analysis were performed on a Hitachi model S-4800 field emission scanning electron microscope (SEM) with an energy-dispersive spectrometer (EDS). Atomic-resolution high-angle annular dark field (HAADF) images were obtained using a spherical-aberration-corrected JEOL ARM200F STEM instrument equipped with double-aberration correctors. TEM samples along the [010] zone-axis direction was prepared through Focused Ion beam (FIB). TEM samples along [001] zone-axis direction was prepared by mechanical exfoliation method with scotch tape. Then resulting suspensions were dispersed onto holey copper grids coated with thin carbon films. Low-temperature magnetization measurements as a function of temperature were performed using a commercial Quantum Design magnetic property measurement system (MPMS), while electrical resistivity was measured with a Quantum Design physical property measurement system (PPMS) by the standard four-probe method.

**Supporting Information**

Supporting information is available from the Wiley Online Library or from the author.

**Conflict of Interest**

Authors declare no conflict of interest.

**Data Availability Statement**

The data that support the findings of this study are available in the Supporting Information of this article.

**Keywords**

Microstructure, Local Structural Distortion, Iron Oligomer, Transmission Electron Microscopy.

**\* Corresponding Author**

hxyang@iphy.ac.cn (Huai-Xin Yang), ljq@aphy.iphy.ac.cn (Jian-Qi Li), junli@iphy.ac.cn (Jun Li)

**Acknowledgments**

This work was supported by the National Natural Science Foundation of China (Grant Nos. U22A6005，12074408, 12074425, 11874422, 52271195), the National Key Research and Development Program of China (Grant Nos. 2021YFA13011502, 2019YFA0308602), the Strategic Priority Research Program (B) of the Chinese Academy of Sciences (Grant Nos. XDB25000000, XDB33000000), the Scientific Instrument Developing Project of the Chinese Academy of Sciences (Grant Nos. YJKYYQ20200055), the Synergetic Extreme Condition User Facility (SECUF), Beijing Municipal Science and Technology major project(Z201100001820006) and IOP Hundred Talents Program (Y9K5051).